\documentclass[graybox]{svmult}

\usepackage{mathptmx}       
\usepackage{helvet}         
\usepackage{courier}        
\usepackage{type1cm}        
\usepackage{makeidx}         
\usepackage{graphicx}        
\usepackage{multicol}        
\usepackage[bottom]{footmisc}

\newtheorem{principle}[theorem]{Principle}

\makeindex             

\begin{document}

\title*{Random Variables Recorded under Mutually Exclusive Conditions: Contextuality-by-Default}
\titlerunning{Contextuality-by-Default}
\author{Ehtibar N. Dzhafarov and Janne V. Kujala}
\authorrunning{E.N. Dzhafarov and J.V. Kujala}
\institute{Ehtibar N. Dzhafarov \at Purdue University, \email{ehtibar@purdue.edu}
\and Janne V. Kujala \at University of Jyv\"askyl\"a, \email{jvk@iki.fi}}
%
%
\maketitle
\vspace{-0ex}
\abstract{We present general principles underlying analysis of the dependence
of random variables (outputs) on deterministic conditions (inputs).
Random outputs recorded under mutually exclusive input values are
labeled by these values and considered stochastically unrelated, possessing
no joint distribution. An input that does not directly influence an
output creates a context for the latter. Any constraint imposed on
the dependence of random outputs on inputs can be characterized by
considering all possible couplings (joint distributions) imposed on
stochastically unrelated outputs. The target application of these
principles is a quantum mechanical system of entangled particles,
with directions of spin measurements chosen for each particle being
inputs and the spins recorded outputs. The sphere of applicability,
however, spans systems across physical, biological, and behavioral
sciences.
}

\section{Introduction}

This paper pertains to any system, physical, biological, or behavioral,
with \emph{random outputs} recorded under varying conditions (\emph{inputs}).
A target example for us is a quantum mechanical system of two entangled
particles, ``Alice's'' and ``Bob's.'' Alice measures the spin
of her particle in one of two directions, $\alpha_{1}$ or $\alpha_{2}$,
and Bob measures the spin of his particle in one of two directions,
$\beta_{1}$ or $\beta_{2}$. Here, $\alpha$ and $\beta$ are inputs,
and each trial is characterized by one of four possible \emph{input
values} $\left(\alpha_{i},\beta_{j}\right)$. The spins recorded in
each trial are realizations of random variables $A$ and $B$, which,
in the simplest case, can attain two values each: $a_{1}$ or $a_{2}$
for $A$ and $b_{1}$ or $b_{2}$ for $B$. One can think of many
examples in other domains with similar formal structure, e.g., a psychophysical
experiment with an observer responding to stimuli with varying characteristics
$\alpha$ (say, intensity) and $\beta$ (say, shape). These characteristics
then constitute inputs, while some characteristics of the responses,
such as response time $A$ (with a continuum of values) and response
correctness $B$ (with two possible values), are random outputs. 

Accounts of the approach presented in this paper can be found in {[}\ref{Dzhafarov,-E.N.,-&2013}-\ref{Dzhafarov,-E.N.,-&New}{]},
but this paper is the first one focusing entirely on its basic principles.
The approach amounts to philosophical rethinking (or at least conceptual
tweaking) of the foundations of probability, specifically, of random
variables and their joint distributions. Here, it is presented without
technical details (that can be reconstructed from {[}\ref{Dzhafarov,-E.N.-&2012a}-\ref{Dzhafarov,-E.N.,-&LNCS13}{]}).

\section{Basic Principles}

Let all or some of the random outputs of a system form a random variable
$X$,%
\footnote{Random variables are understood in the broadest sense, so that a vector
of random variables (or any set thereof, or a random process) is a
random variable too.%
} and the totality of all inputs be a variable $\chi$. In our target
example, $\chi=\left(\alpha,\beta\right)$ with input values $\chi_{1}=\left(\alpha_{1},\beta_{1}\right)$,
$\ldots$, $\chi_{4}=\left(\alpha_{2},\beta_{2}\right)$, whereas
$X$ can be $\left(A,B\right)$ with values $x_{1}=\left(a_{1},b_{1}\right)$,
$\ldots$, $x_{4}=\left(a_{2},b_{2}\right)$, or $A$ with values
$a_{1},a_{2}$, or $B$ with values $b_{1},b_{2}$. If $\chi$ itself
is a random variable, so that $\chi_{1},\chi_{2},\ldots$ occur with
some probabilities, we ignore these probabilities and simply condition
the recorded outputs $X$ on values of $\chi$. In other words, we
have a distribution of $X$ given that $\chi=\chi_{1}$, a distribution
of $X$ given that $\chi=\chi_{2}$, etc., irrespective of whether
we can control and predict the values of $\chi$, or they occur randomly.
Now, this conditioning upon input values means that $X$ \emph{is
indexed by different values of} $\chi$. We obtain thus, ``automatically,''
a set of different random variables in place of what we previously
called a random variable $X$. We have $X_{\chi_{1}}$ (or $X_{1}$,
if no confusion is likely) which is $X$ when $\chi=\chi_{1}$, $X_{\chi_{2}}$
(or $X_{2}$) which is $X$ when $\chi=\chi_{2}$, etc. Let us formulate
this simple observation as a formal principle.

\begin{principle}Outputs recorded under different (hence mutually
exclusive) input values are labeled by these input values and considered
different random variables. These random variables are stochastically
unrelated, i.e., they possess no joint distribution.\end{principle}

Thus, in our target example, we have four random variables $A_{ij}$,
four random variables $B_{ij}$, and four random variables $\left(A,B\right)_{ij}=\left(A_{ij},B_{ij}\right)$
corresponding to the four input values $\chi_{k}=\left(\alpha_{i},\beta_{j}\right)$.
The principle holds irrespective of how the distribution of $X_{k}$
depends on $\chi_{k}$. Thus, the variables $A_{i1}$ and $A_{i2}$
remain different even if their distributions are identical (as they
should be if Bob's choice cannot influence Alice's measurements).
One must not assume that they are one and the same random variable,
$A_{i}=A_{i1}=A_{i2}$. The latter would mean that $A_{i1}$ and $A_{i2}$
have a joint distribution, because of which the probabilities $\Pr\left[A_{i1}=A_{i2}\right]$
are well defined, and that these probabilities equal 1. But $A_{i1}$
and $A_{i2}$ do not have a joint distribution. Indeed, two random
variables $X$ and $Y$ have a joint distribution only if their values
can be thought of as observed ``in pairs,'' i.e., if there is a
scheme of establishing correspondence $x_{\left(i\right)}\leftrightarrow y_{\left(i\right)}$
between observations $x_{\left(1\right)},x_{\left(2\right)},\ldots$
of $X$ and $y_{\left(1\right)},y_{\left(2\right)},\ldots$ of $Y$.
In our example, the correspondence is defined by the two measurements
being simultaneously performed on a given pair of entangled particles.
Each such a pair of measurements corresponds to a certain input value,
e.g., $A_{21}$ and $B_{21}$ correspond to $\chi=\left(\alpha_{2},\beta_{1}\right)$.
Therefore, no measurement outputs corresponding to different input
values, such as $A_{i1}$ and $A_{i2}$, or $A_{i1}$ and $B_{i2}$
co-occur in the same sense in which, say, $A_{i1}$ co-occurs with
$B_{i1}$.

However, given any two random variables $X$ and $Y$, one can impose
on them a joint distribution, and create thereby a random variable
$Z=\left(X,Y\right)$, referred to as a \emph{coupling} for $X$ and
$Y$. By definition, the distribution of a coupling $Z$ agrees with
the distributions of $X$ and $Y$ as its marginals. 

\begin{principle}Stochastically unrelated outputs recorded under
mutually exclusive input values can be coupled (imposed a joint distribution
upon) arbitrarily. There are no privileged couplings.\end{principle}

Thus, in our target example, the famous Bell-type theorems {[}\ref{Bell,-J.:-On},\ref{Dzhafarov,-E.N.-&2012a},\ref{Fine,-A.:-Hidden}{]}
implicitly impose on $\left(A_{11},B_{11}\right)$, $\ldots$, $\left(A_{22},B_{22}\right)$
a coupling with $A_{i1}=A_{i2}$ and $B_{1j}=B_{2j}$. This amounts
to considering a random variable $\left(A'_{1},A'_{2},B'_{1},B'_{2}\right)$
such that $\left(A'_{i},B'_{j}\right)$ is distributed as $\left(A_{ij},B_{ij}\right)$.
The Bell-type theorems show that such a coupling exists if and only
if the distributions of the coupled pairs $\left(A_{11},B_{11}\right)$,
$\ldots$, $\left(A_{22},B_{22}\right)$ satisfy certain constraints
(Bell-type inequalities, known to be violated in quantum mechanics).
In our approach, however, except possibly for simplicity considerations,
this coupling has no privileged status among all possible coupling
for $\left(A_{11},B_{11}\right)$, $\ldots$, $\left(A_{22},B_{22}\right)$.
Thus, any distribution of spins satisfying Bell-type inequalities is also compatible with the coupling
in which $\left(A_{11},B_{11}\right)$, $\ldots$, $\left(A_{22},B_{22}\right)$
are stochastically independent pairs of random variables, as well
as with an infinity of other couplings in which $\Pr\left[A_{i1}=A_{i2}\right]$
and $\Pr\left[B_{1j}=B_{2j}\right]$ may be different from 1. 

If the distributions of $A_{i1}$ and $A_{i2}$ are not the same for
$i=1$ or $i=2$, the situation is simple: the output $A$ is influenced
by both inputs $\alpha$ and $\beta$ (and analogously for $B_{1j}$
and $B_{2j}$). If, however, the distributions of $A_{i1}$ and $A_{i2}$
are always the same, and if, moreover, substantive considerations
(e.g., laws of special relativity) prevent the possibility of interpreting
$\beta$ as ``directly'' influencing $A$, then we can say that
$\beta$ forms a \emph{context} for the dependence of $A$ on $\alpha$
(and analogously for $\alpha$ creating a context for the dependence
of $B$ on $\beta$). Principle 1 ensures that this contextuality
is introduced ``automatically,'' by labeling all outputs by all
conditions under which they are recorded. The degree and form of contextuality
in a given system (e.g., those with constraints more relaxed than
the Bell-type inequalities {[}\ref{Cirel'son,-B.S.:-Quantum},\ref{Landau,-L.-J.:}{]})
can be characterized by considering all possible probabilities $\Pr\left[A_{i1}=A_{i2}\right]$
and $\Pr\left[B_{1j}=B_{2j}\right]$, called \emph{connection probabilities}
in {[}\ref{Dzhafarov,-E.N.,-&2013}-\ref{Dzhafarov,-E.N.,-&New}{]}.
This approach allows one to embark on a deeper investigation of the
relationship between the classical probability theory and quantum
mechanics than in the Bell-type theorems.

\section{Apparent Problems with the Approach}

Two objections can be raised against our approach. One is that it
requires to label random variables by circumstances that cannot possibly
be relevant. If reaction time $X$ to a given stimulus is recorded
in conjunction with measurements of the temperature on Mars with the
values $\chi_{1}=low$ and $\chi_{2}=high$, would it be meaningful
to ``automatically'' split $X$ into stochastically unrelated $X_{low}$
and $X_{high}$? The answer is: it is meaningful. If the temperature
on Mars affects the distribution of $X$, then considering $X_{low}$
and $X_{high}$ as different random variables is clearly useful for
understanding of $X$. If, as we suspect, the temperature on Mars
does not affect the distribution of $X$, then one can impose on $\left(X_{low},X_{high}\right)$
an arbitrary coupling, including one with $X_{low}=X_{high}=X$. The
latter choice amounts to ignoring the temperature on Mars altogether.

The other objection is that if we apply Principle 1 systematically,
we have to consider different realizations of a random variable $X$
as stochastically unrelated random variables. $X$ occurring in trial
1 as $x_{(1)}$ is labeled $X_{1}$ and considered stochastically unrelated
to $X_{2}$ that occurs in trial 2 as $x_{(2)}$, and so on. But this is perfectly reasonable, and moreover, it is a standard
issue in the probabilistic theory of couplings {[}6{]}. Once a coupling
(e.g., the commonly used iid one) is imposed on $X_{1},X_{2},\ldots$,
it creates a new random variable $Y=\left(X_{1},X_{2},\ldots\right)$,
of which we have a single realization $y=\left(x_{(1)},x_{(2)},\ldots\right)$.
One can then investigate whether this $y$ is statistically plausible
in view of the distribution of $Y$ using standard statistical reasoning.

\end{document}